\documentclass[runningheads]{llncs}
\usepackage[T1]{fontenc}
\usepackage{graphicx}
\usepackage{amsmath}
\usepackage{amsfonts}
\usepackage{mathtools}
\usepackage[ruled]{algorithm2e}
\usepackage{float}
\usepackage{tikz}
\usetikzlibrary{positioning}
\usetikzlibrary{calc}
\usepackage{booktabs}       
\usepackage{multirow}
\usepackage[caption=false]{subfig}

\DeclarePairedDelimiter\norm{\lVert}{\rVert}%
\DeclareMathOperator{\st}{s.t.}
\DeclareMathOperator*{\argmin}{arg\,min}

\usepackage[normalem]{ulem}
\usepackage{xcolor}

\def\myFigureScale{0.115}
\def\mycolumnsep{0.5}
\def\myrowsep{2.6}

\begin{document}
\title{Joint Manifold Learning and Optimal Transport for Dynamic Imaging}
\titlerunning{Joint Manifold Learning and Optimal Transport for Dynamic Imaging}
%
\author{Sven Dummer \inst{1}\orcidID{0000-1111-2222-3333} \and
Christoph Brune \inst{2,3}\orcidID{1111-2222-3333-4444} \and
Third Author\inst{3}\orcidID{2222--3333-4444-5555}}
\author{Sven Dummer \orcidID{0000-0002-5276-4155}, Puru Vaish 
\orcidID{0000-0002-5180-5293} \and
Christoph Brune \orcidID{0000-0003-0145-5069}}
\authorrunning{S. Dummer, P. Vaish, and C. Brune}
%

\institute{University of Twente, Enschede, Overijssel, 7522 NB, the Netherlands \\
\email{\{s.c.dummer, p.vaish, c.brune\}@utwente.nl}}

\maketitle              
\begin{abstract}
Dynamic imaging is critical for understanding and visualizing dynamic biological processes in medicine and cell biology. These applications often encounter the challenge of a limited amount of time series data and time points, which hinders learning meaningful patterns. Regularization methods provide valuable prior knowledge to address this challenge, enabling the extraction of relevant information despite the scarcity of time-series data and time points. In particular, low-dimensionality assumptions on the image manifold address sample scarcity, while time progression models, such as optimal transport (OT), provide priors on image development to mitigate the lack of time points. Existing approaches using low-dimensionality assumptions disregard a temporal prior but leverage information from multiple time series. OT-prior methods, however, incorporate the temporal prior but regularize only individual time series, ignoring information from other time series of the same image modality. In this work, we investigate the effect of integrating a low-dimensionality assumption of the underlying image manifold with an OT regularizer for time-evolving images. In particular, we propose a latent model representation of the underlying image manifold and promote consistency between this representation, the time series data, and the OT prior on the time-evolving images. We discuss the advantages of enriching OT interpolations with latent models and integrating OT priors into latent models. 
\end{abstract}

\section{Introduction}
Dynamic imaging is crucial in many fields, particularly medical imaging and cell biology. It enables visualization and spatiotemporal modeling of time-dependent biological processes, offering a deeper understanding of biological processes and dynamics. These insights help experts to better understand and potentially influence how cells and diseases develop over time.

Dynamic imaging models for visualization and understanding require data, which is often scarce. The number of image time series is generally small due to a limited number of patients or cells. Moreover, these time series typically contain relatively few data points over time. This scarce data makes it challenging to develop accurate dynamic imaging models that generalize well across cases and capture the underlying trends of disease progression or cell development. 

The small data scenario requires prior knowledge about the underlying image manifold and the plausible development of the images. One well-known prior assumption on the image manifold is its relatively low dimensionality. State-of-the-art models in fields such as generative modeling \cite{rombach2022high} and inverse problems \cite{song2024solving} utilize this prior and learn representations of plausible images via low-dimensional latent vectors. 

Another important prior is Optimal Transport (OT). OT can be used as prior knowledge on the underlying image manifold via a patch prior \cite{hertrich2022WassersteinPatch} or via a template-based prior \cite{gao2023template}. For the time-related data scarcity, OT provides priors for image development via the dynamical Benamou-Brenier formulation \cite{bredies2023generalized,maas2015generalized,bon2024optimal,brune20104d,chen2021spatiotemporal}.

\begin{figure}
    \centering
    \includegraphics[width=0.85\linewidth]{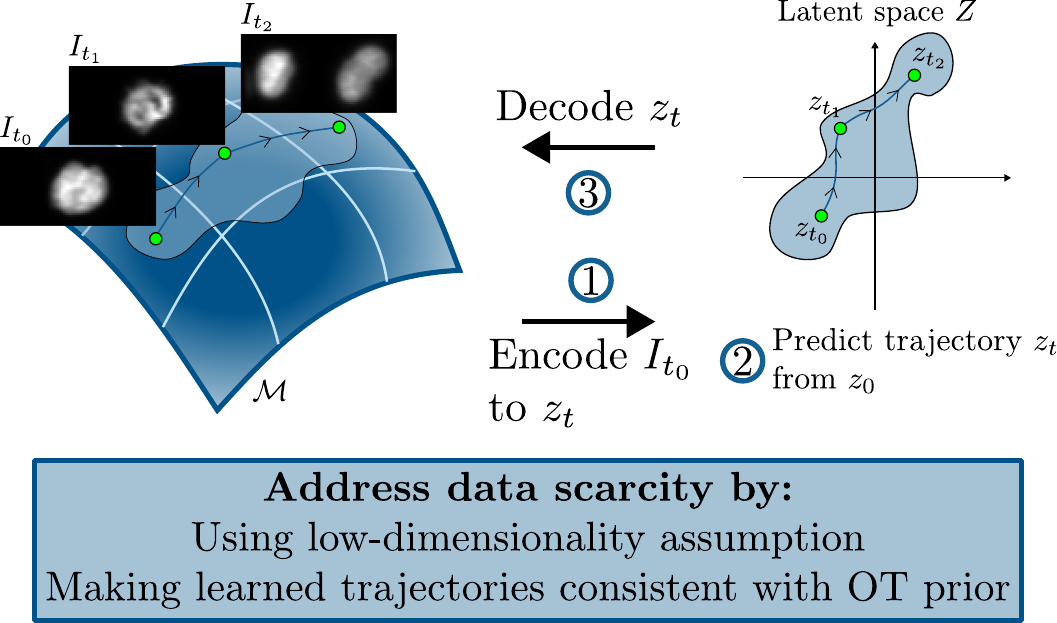}
    \caption{\textbf{Model overview.} An initial image is encoded into latent space, where a trajectory is predicted and decoded back into image space. To address the scarcity of time series data and time points, existing methods often rely on priors such as a low-dimensional manifold assumption or a dynamic OT prior. Our model combines these approaches by constructing a low-dimensional latent space and regularizing the predicted image trajectories to align with the OT prior.}
    \label{fig:intro_picture}
\end{figure}
While existing approaches based on the low-dimensionality assumption analyze the data without accounting for such temporal priors, methods leveraging OT priors focus solely on regularizing individual time series, neglecting information from other time series lying on the same data manifold. As depicted in Figure \ref{fig:intro_picture}, this work explores combining the low-dimensionality assumption with a temporal OT prior. In particular, we introduce a latent model that uses the dynamic OT prior and is applicable to dynamic imaging. Using an autoencoder as a (learned) low-dimensional latent representation of plausible images, we map predicted latent-space trajectories to image-space trajectories, which are regularized via the dynamical OT prior. We use this model and cell biology-inspired experiments to examine how combining latent space models and dynamical optimal transport priors affects manifold learning of dynamic and static images and how it affects image interpolation and extrapolation.

\section{Related Works}

\subsection{Dynamic Imaging via Optimal Transport, LDDMM, and Metamorphosis}
\label{sec:rel_work_dyn_img}

Optimal transport has been used as template-based regularization \cite{gao2023template} for static imaging and as a regularizer of temporal dynamics in dynamic imaging \cite{bredies2023generalized,maas2015generalized,brune20104d,chen2021spatiotemporal,boink2016combined}. Similar models, such as Large Deformation Diffeomorphic Metric Mapping (LDDMM) and metamorphosis, have also been applied as template-based regularizer \cite{lang2020template,chen2018indirect,neumayer2019regularization,gris2020image,neumayer2024template} and time dynamic regularizer \cite{burger2018variational,chen2019new,hauptmann2021image}. 

Dynamic imaging methods that employ OT, LDDMM, or metamorphosis typically focus on a single dynamic image and use an additional predefined static image prior, such as total variation (TV) or a segmentation mask that constrains the location of non-zero pixel intensities \cite{boink2016combined}. These priors constrain the OT, LDDMM, or metamorphosis geodesics, which are used to regularize temporal dynamics. 

In contrast, our approach introduces a static image manifold prior, where the manifold of plausible (static) images is not predefined but learned from multiple dynamic images. Our model shares similarities with a recent model \cite{dummer2023rda} that combines latent modeling and LDDMM to regularize models for joint shape latent modeling and registration. However, their approach focuses on static objects, adopting a template-based regularization viewpoint, whereas our method applies to dynamic objects.

\subsection{Dimensionality Reduction in Wasserstein Space}
Combining OT and dimensionality reduction has recently gained considerable interest. For instance, one approach learns a dictionary within Wasserstein space \cite{schmitz2018wasserstein}, while another \cite{hamm2023wassmap} extends Isomap to this setting. The differential geometric structure of (learned) manifolds in Wasserstein space has also been analyzed mathematically \cite{hamm2023manifold}. In the context of latent space models for images, OT-based methods have been used to enhance latent interpolation. For instance, a recent approach \cite{simon2020barycenters} applies an OT-based scheme after the model has been trained on the data, while another \cite{feng2024improving} adds a regularization term that promotes linear latent interpolations to follow OT geodesics. While these works explore combining autoencoders and OT, they do not address the integration of time series data and OT geodesics.

\section{Dynamical Optimal Transport and Barycenters}
Optimal transport is defined as optimally moving a probability distribution to another probability distribution.
\begin{definition}[Wasserstein space $\mathcal{W}_2(\mathbb{R}^d)$ \cite{villani2009optimal}]
Assume we have $\mu, \nu \in P(\mathbb{R}^d)$ with $P(\mathbb{R}^d)$ the space of Borel probability measures. For an arbitrary $x_0 \in \mathbb{R}^d$, $(\mathcal{W}_p(\mathbb{R}^d), W_p)$ is the $p$-Wasserstein space with:
\begin{equation*}
    \begin{aligned}
        W_p(\mu, \nu) := \left(\inf_{\pi \in \Pi(\mu, \nu)} \int_{\mathbb{R}^d} \norm{x - y}_p ^p d\pi(x,y)\right)^\frac{1}{p} \\
        \mathcal{W}_p(\mathbb{R}^d):= \{\mu \in P(\mathbb{R}^d); \quad \int_{\mathbb{R}^d} \norm{x_0 -x}_p^p d\mu(x) < \infty\}
    \end{aligned}
\end{equation*}
where $\Pi(\mu, \nu) := \{\pi \in P(\mathbb{R}^d \times \mathbb{R}^d); \int_{\mathbb{R}^d} \pi(dx, y) = \nu, \int_{\mathbb{R}^d} \pi(x, dy) = \mu \}$.
\end{definition}
For $p=2$, the Wasserstein distance has a time-dependent reformulation.
\begin{theorem}[\cite{Ambrosio2013,ambrosio2005gradient}]
    The $2$-Wasserstein distance can be rewritten as:
    \begin{equation*}
        W_2(\mu_0, \mu_1) = \inf_{\mu_t} \int_0^1 |\dot{\mu}_t| dt
    \end{equation*}
    where the infimum is over absolutely continuous curves with starting point $\mu_0$, endpoint $\mu_1$, and $|\dot{\mu}_t|$ the metric derivative of the curve $\mu_t$:
    \begin{equation*}
        |\dot{\mu}| := \lim_{h \rightarrow 0}\frac{W_2(\mu_{t+h}, \mu_t)}{|h|}.
    \end{equation*}
    \label{thm:dyn_formulation_abs_cont_curves}
\end{theorem}
The reformulation resembles a Riemannian geodesic and yields a weak Riemannian structure on $\mathcal{W}_2$ via definitions of tangent vectors and Riemannian metrics  \cite{Ambrosio2013,ambrosio2005gradient,otto2001geometry}. Instead of this time-dependent formulation of Riemannian OT geodesics, we can alternatively define OT geodesics via the concept of Wasserstein barycenters.
\begin{definition}[Wasserstein barycenters and interpolations]
    Let $\mu_i \in P(\mathbb{R}^d)$ for $i\in\{1, \ldots, n\}$. Then for some sequence $\alpha_i$ with $\sum_{i=1}^n \alpha_i = 1$, a Wasserstein barycenter is defined as:
    \begin{equation*}
        \mu := \argmin_{\mu \in P(\mathbb{R}^d)} \left(\sum_{i=1}^n \alpha_i W_2(\mu_i, \mu)\right).
    \end{equation*}
    In particular, for $t \in [0, 1]$, we define the barycentric interpolation as:
    \begin{equation*}
        B(\mu_1, \mu_2, t):=\argmin_{\mu \in P(\mathbb{R}^d)} \left((1-t) W_2(\mu_1, \mu) + t W_2(\mu_2, \mu)\right) = \mu_t
    \end{equation*}
    with $\mu_t$ the geodesic defined in Theorem \ref{thm:dyn_formulation_abs_cont_curves}.
    \label{def:wass_barycenter}
\end{definition}

\section{Autoencoder Model for Dynamic imaging}
As shown in Figure \ref{fig:intro_picture}, our approach uses an autoencoder as a low-dimensional manifold prior and combines this prior with dynamical OT regularization. Given $N$ time series $\{I^{i}_j\}_{j=1}^{N_t}$ of length $N_t$, indexed by $i \in \{1, \ldots, N\}$, our model first encodes each time point $j$ in time series $i$ into a latent code $z_{ij}=E(I_j^i)$, where the encoder $E\colon \mathbb{R}^m \rightarrow Z$ maps images to a latent space $Z=\mathbb{R}^d$. 

To model temporal evolution, we then evolve the initial latent codes $z_{i1}$ via a neural ordinary differential equation in latent space. In particular, given a neural network $v_\varphi \colon Z \to Z$ parameterizing the right-hand side of the neural ODE, we evolve the initial latent codes $z_{i1}$ via:
\begin{equation*}
    \frac{\mathrm{d}z^i(t)}{\mathrm{d}t} = v_\varphi(z^i(t)), \quad z^i(0) = z_1^i.
\end{equation*}
Since $v_\varphi$ depends only on the current state, supplying an initial condition determines the future trajectory, enabling image prediction. 

Subsequently, a decoder $D \colon Z \rightarrow \mathbb{R}^m$ generates images $D(z_{ij})$ and $D(z^i(t_j))$ using the static and dynamic encodings, respectively. Both reconstructions should match the ground truth data $I_j^i$. At the same time, the latent codes $z_{ij}$ and $z^i(t_j)$ should be consistent since they represent the same image $I_j^i$. Our model ensures accurate reconstructions and latent consistency via the loss function:
\begin{equation*}
    \norm{D(z^i(t_j)) - I_j^i}_2^2 + \gamma_1 \norm{D(z_j^i) - I_j^i}_2^2 + \gamma_2 \norm{z^i(t_j) - z_j^i}_2^2,
\end{equation*}
where $\gamma_1,\gamma_2 \in \mathbb{R}$.

Finally, we introduce a Riemannian dynamical OT regularizer, which imposes an OT Riemannian structure on the learned manifold. This regularizer encourages the reconstructed time series, once normalized, to approximate OT geodesics. In particular, using $n(I) := I / \int_\Omega I(x) \mathrm{d}x$ as normalization function, the normalized reconstruction $\mu^i_{t}:=n(D(z^i(t)))$ for $t_j < t < t_{j+1}$ should be close to the Wasserstein barycentric interpolation $B\left(\mu^i_{t_j}, \mu^i_{t_{j+1}}, \frac{t-t_j}{t_{j+1}-t_j}\right)$, as defined in Definition \ref{def:wass_barycenter}. The normalization is required since OT assumes unit mass distributions. However, as the decoder does not enforce the unit mass constraint, this leaves a degree of freedom in the mass of the image. To address this, we encourage a linear interpolation between the masses of the boundary images using:
\begin{equation*}
    s^i_j(t)=\left(1 - \frac{t- t_{j}}{t_{j+1}-t_{j}}\right)\norm{D(z^i(t_j))}_1 + \left(\frac{t - t_{j}}{t_{j+1}-t_{j}}\right)\norm{D(z^i(t_{j+1}))}_1,
\end{equation*}
In particular, the Riemannian dynamical OT regularizer enforces proximity to the scaled barycentric interpolation:
\begin{equation*}
    \left \lVert D(z^i(t)) -  s^i_j(t)B\left(\mu^i_{t_j}, \mu^i_{t_{j+1}}, \frac{t- t_{j}}{t_{j+1}-t_{j}}\right)\right\rVert_2^2
\end{equation*}

Bringing everything together, our model corresponding to Figure \ref{fig:intro_picture} solves the following optimization problem:
\begin{equation*}
    \begin{aligned}
    &\min \limits_{E, D, v_\varphi} && \frac{1}{N \cdot N_t}\sum_{i=1}^{N} \sum_{j=1}^{N_t}\left( \norm{D(z^i(t_j)) - I_j^i}_2^2 + \gamma_1 \norm{D(z_j^i) - I_j^i}_2^2 + \gamma_2 \norm{z^i(t_j) - z_j^i}_2^2\right) \\ 
    &&& + \lambda \sum_{i=1}^N \sum_{j=1}^{N_t-1} \int_{t_j}^{t_{j+1}} \left \lVert D(z^i(t)) -  s^i_j(t)B\left(\mu^i_{t_j}, \mu^i_{t_{j+1}}, \frac{t- t_{j}}{t_{j+1}-t_{j}}\right)\right\rVert_2^2 \mathrm{d}t\\
    &\st && z_j^i = E(I_j^i), \\
    & && \frac{\mathrm{d}z^i(t)}{\mathrm{d}t} = v_\varphi(z^i(t)), \quad z^i(0) = z_1^i, \\
    & && \mu^i_t = n(D(z^i(t))), \\
    & && s^i_j(t)=\left(1 - \frac{t- t_{j}}{t_{j+1}-t_{j}}\right)\norm{D(z^i(t_j))}_1 + \left(\frac{t - t_{j}}{t_{j+1}-t_{j}}\right)\norm{D(z^i(t_{j+1}))}_1,
    \end{aligned}
\end{equation*}

\section{Numerical Experiments}
This section presents numerical examples\footnote{Our code is available at: \url{https://github.com/SCdummer/joint-manifold-learning-and-ot}} to illustrate the impact of joining manifold learning and dynamic OT. In particular, we test the OT regularizer against an $l_2$ regularizer $\norm{\frac{\mathrm{d}}{\mathrm{d}t}D(z(t))}_2$ on the change in the image and an $l_2$ regularizer$\norm{\frac{\mathrm{d}}{\mathrm{d}t}z(t)}_2$ on the change in the latent code. We focus on these regularizers because more complex ones, such as brightness consistency in optical flow, require solving an additional variational problem, making efficient backpropagation challenging. While Wasserstein regularization also imposes a constraint in the form of mass conservation, it benefits from efficient entropy-regularized Sinkhorn solvers and their differentiation properties, enabling efficient integration into our model. Besides investigating the benefit of OT regularization, we also investigate the benefit of the learned manifold by comparing our neural ODE prediction model to simple $l_2$ and Wasserstein interpolations of the images. 

We use a synthetic Gaussian dataset and the HeLa cells dataset from the Cell Tracking Challenge \cite{ulman2017objective,mavska2023cell} for the experiments. The synthetic Gaussian dataset consists of image time series depicting Gaussians that move from the left to the right while growing in size. The movement and growth accelerate in the middle, simulating a discontinuous transition from one state to another when the time series has insufficient temporal resolution. The HeLa dataset contains time series that transition from a single cell to two cells, presenting a challenge similar to, but slightly more complex than, the synthetic Gaussians dataset.

\subsection{Effect of OT Regularization on Manifold-Based Interpolation} \label{sec:frame_interpolation}
To investigate the impact of the OT regularizer, we first explore its effect on image interpolations on the learned manifold. We train the models on time series whose data is subsampled every $5$ time steps. This subsampling ensures missing images and can be used to assess the quality of the interpolations. 

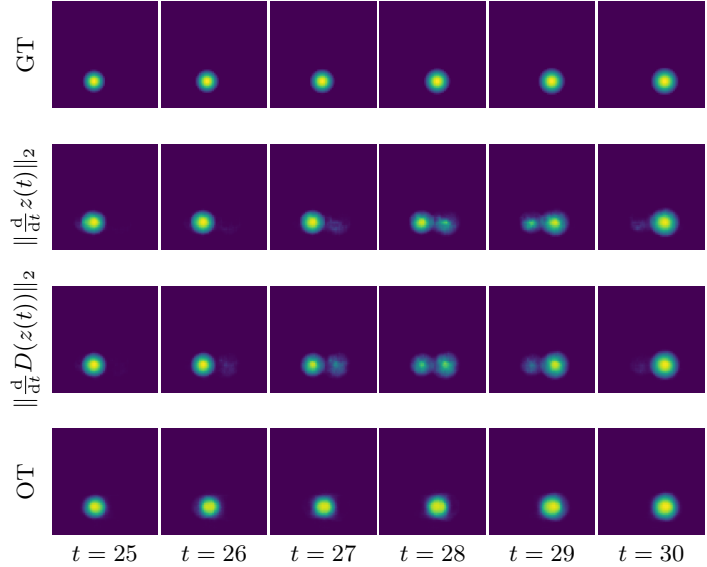
\begin{figure}[!b]
    \begin{center}
    \begin{tikzpicture}[
 image/.style = {text width=\myFigureScale\textwidth, 
                 inner sep=0pt, outer sep=0pt},
node distance = \myrowsep mm and \mycolumnsep mm
                        ] 

\matrix[column sep=\mycolumnsep mm, row sep=3mm] (gt_matrix) {
    \node [image] (frame1_gt) {\includegraphics[width=\linewidth]{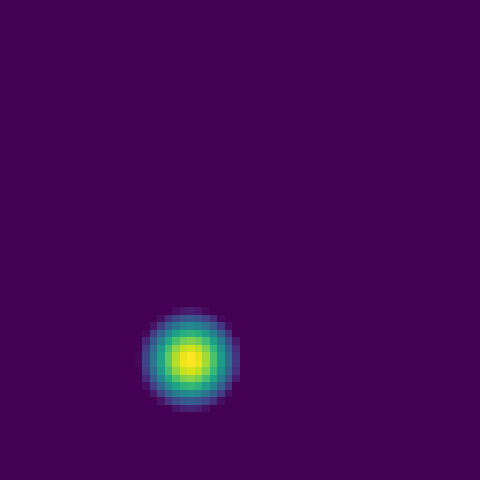}}; & 
    \node [image] {\includegraphics[width=\linewidth]{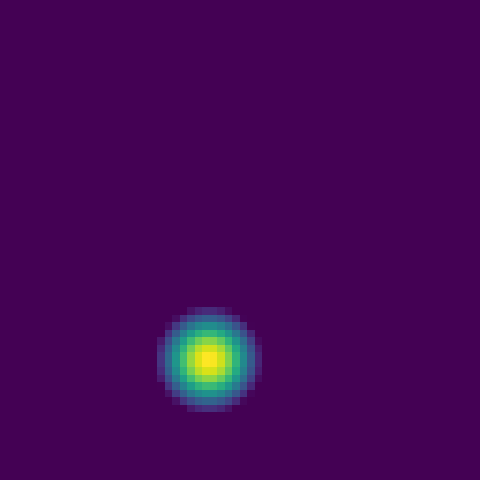}}; & 
    \node [image] {\includegraphics[width=\linewidth]{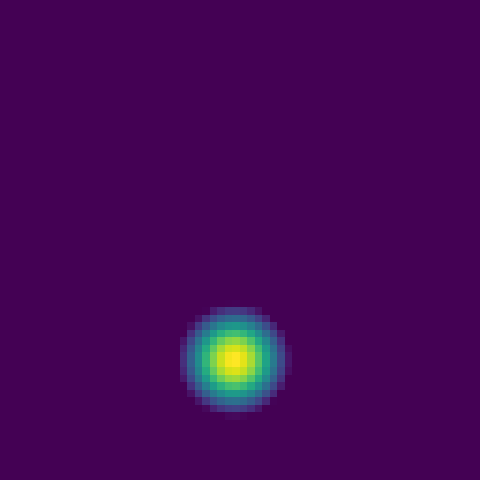}}; & 
    \node [image] {\includegraphics[width=\linewidth]{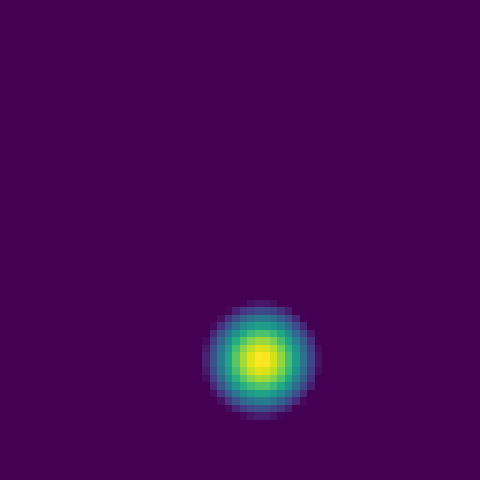}}; & 
    \node [image] {\includegraphics[width=\linewidth]{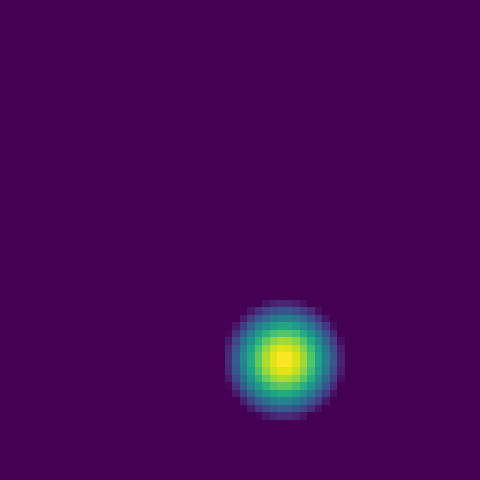}}; &
    \node [image] {\includegraphics[width=\linewidth]{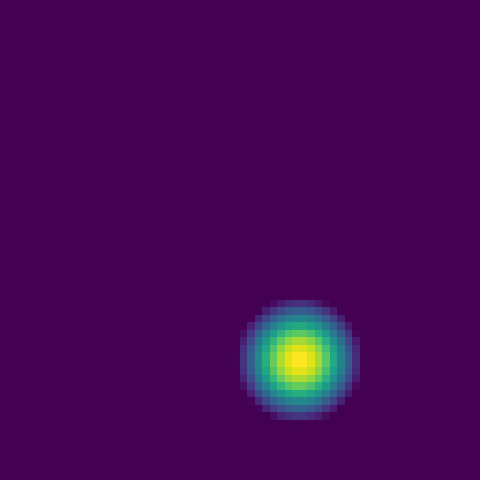}}; \\
};


\matrix[below=of gt_matrix, column sep=\mycolumnsep mm, row sep=3mm] (matrix_dzdt) {
    \node [image] (frame1_dzdt) {\includegraphics[width=\linewidth]{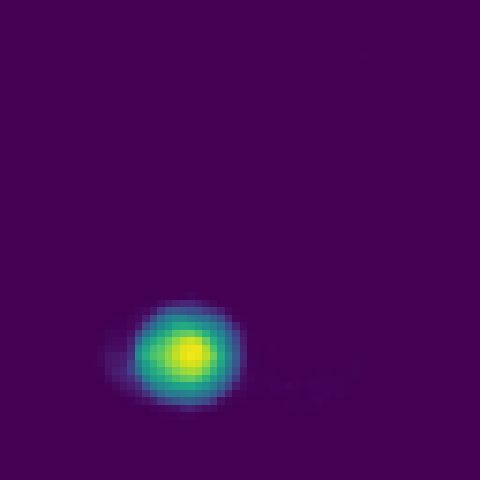}}; & 
    \node [image] {\includegraphics[width=\linewidth]{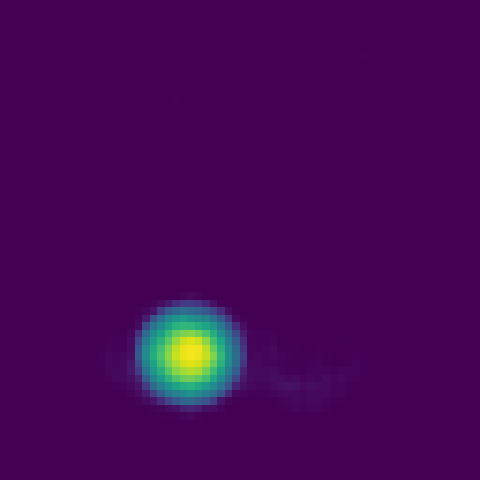}}; & 
    \node [image] {\includegraphics[width=\linewidth]{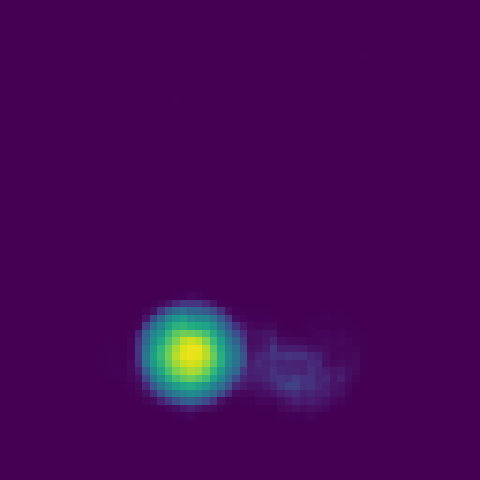}}; & 
    \node [image] {\includegraphics[width=\linewidth]{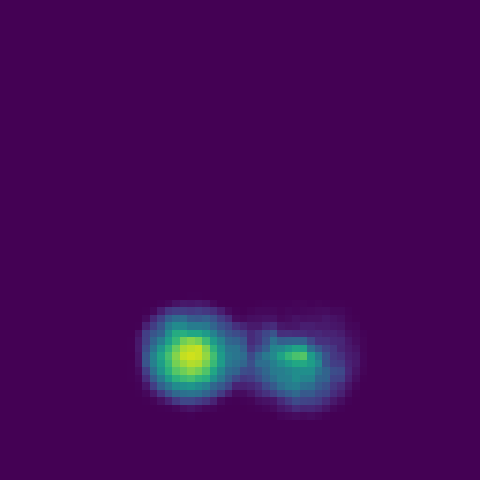}}; &
    \node [image] {\includegraphics[width=\linewidth]{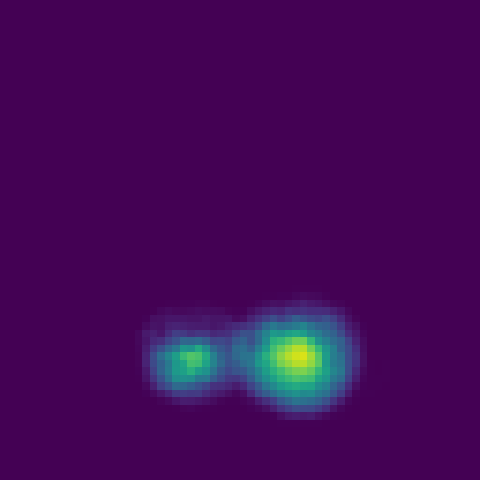}}; &
    \node [image] {\includegraphics[width=\linewidth]{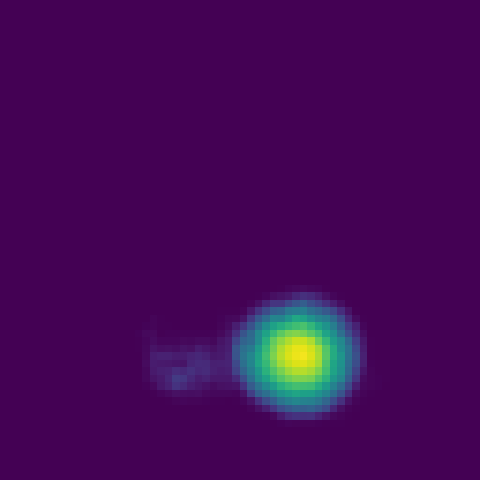}}; \\ 
};

\matrix[below=of matrix_dzdt, column sep=\mycolumnsep mm, row sep=3mm] (matrix_dIdt){
    \node [image] (frame1_dIdt) {\includegraphics[width=\linewidth]{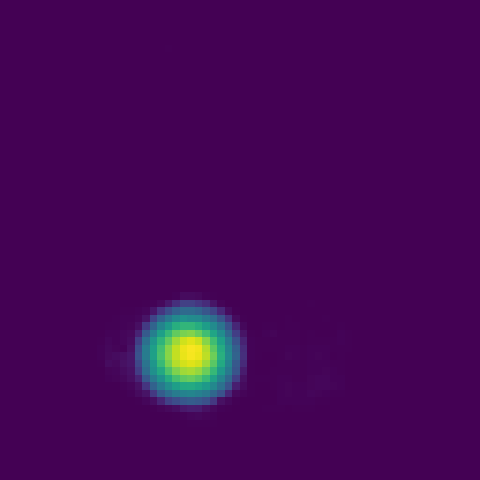}}; & 
    \node [image] {\includegraphics[width=\linewidth]{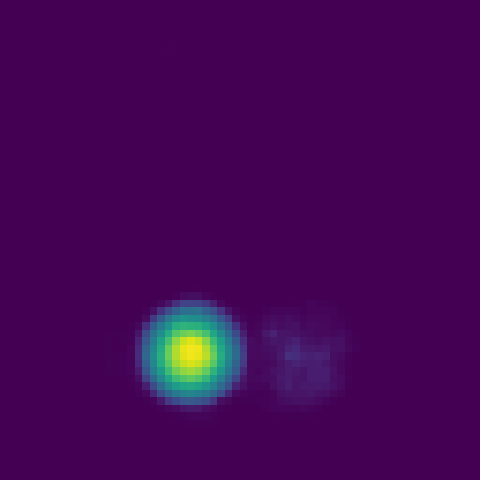}}; & 
    \node [image] {\includegraphics[width=\linewidth]{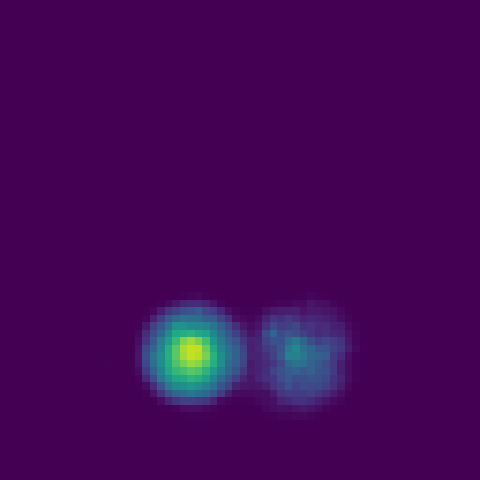}}; & 
    \node [image] {\includegraphics[width=\linewidth]{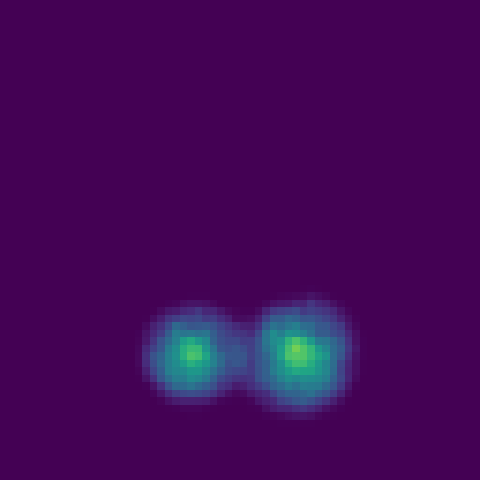}}; &
    \node [image] {\includegraphics[width=\linewidth]{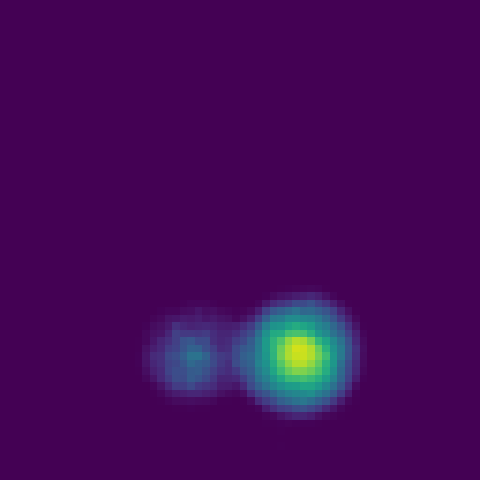}}; &
    \node [image] {\includegraphics[width=\linewidth]{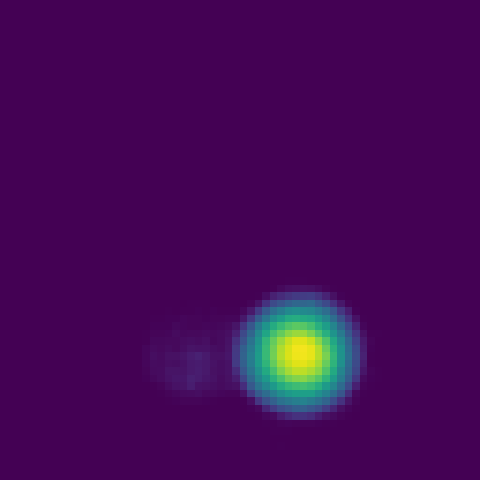}}; \\ 
};

\matrix[below=of matrix_dIdt, column sep=\mycolumnsep mm, row sep=3mm] (matrix_OT) {
    \node [image] (frame1) {\includegraphics[width=\linewidth]{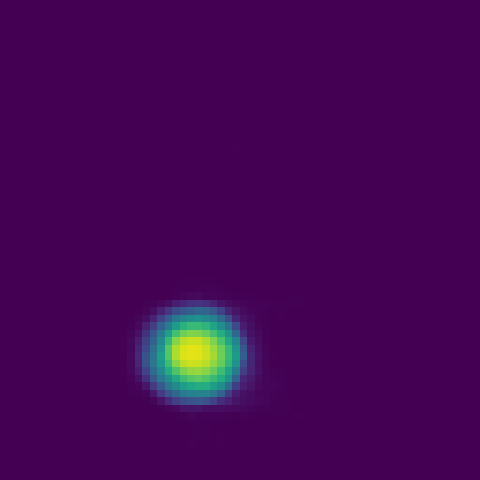}}; & 
    \node [image] (frame2) {\includegraphics[width=\linewidth]{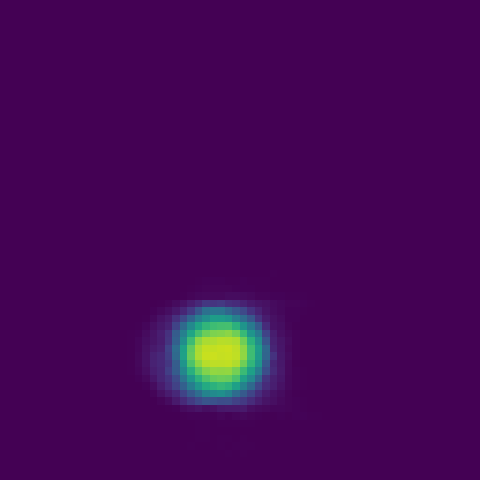}}; & 
    \node [image] (frame3) {\includegraphics[width=\linewidth]{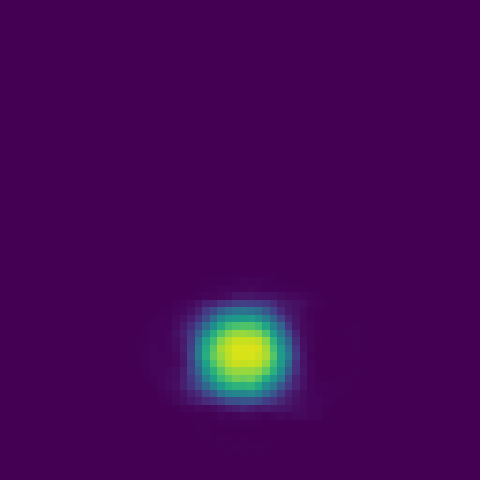}}; & 
    \node [image] (frame4) {\includegraphics[width=\linewidth]{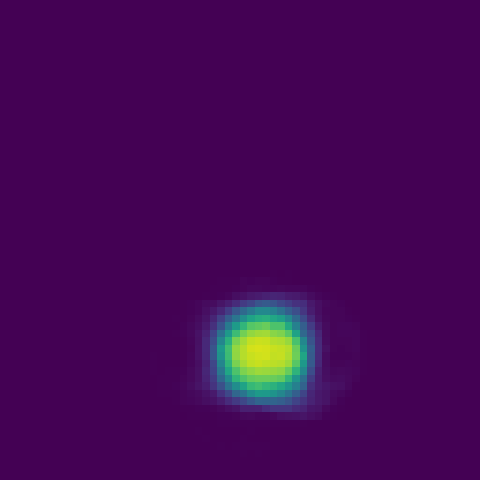}}; &
    \node [image] (frame5) {\includegraphics[width=\linewidth]{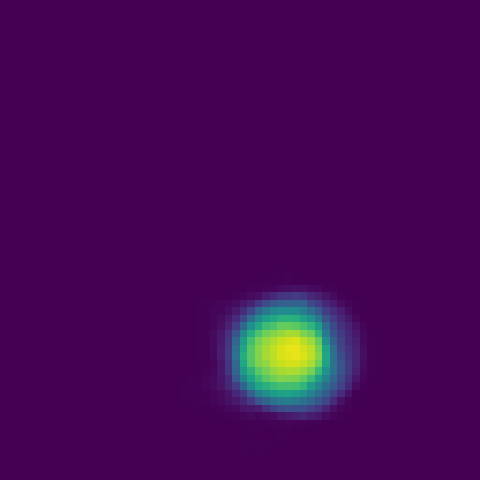}}; &
    \node [image] (frame6) {\includegraphics[width=\linewidth]{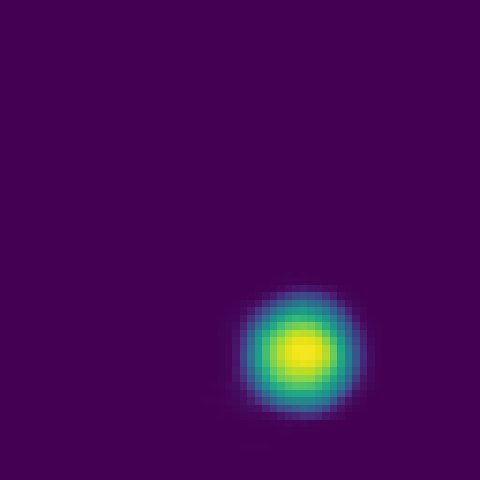}}; \\ 
};

\node [left=of frame1_dIdt, rotate=90, anchor=south] (dI_dt) {$\norm{\frac{\mathrm{d}}{ \mathrm{d}t}D(z(t))}_2$};
\draw (dI_dt |- gt_matrix) node[rotate=90] {GT};
\draw (dI_dt |- matrix_dzdt) node[rotate=90] {$\norm{\frac{\mathrm{d}}{\mathrm{d}t}z(t)}_2$};
\draw (dI_dt |- matrix_OT) node[rotate=90] {OT};

\node at ($(frame1.south) + (0.0,-0.25)$) {$t=25$};
\node at ($(frame2.south) + (0.0,-0.25)$) {$t=26$};
\node at ($(frame3.south) + (0.0,-0.25)$) {$t=27$};
\node at ($(frame4.south) + (0.0,-0.25)$) {$t=28$};
\node at ($(frame5.south) + (0.0,-0.25)$) {$t=29$};
\node at ($(frame6.south) + (0.0,-0.25)$) {$t=30$};

\end{tikzpicture}
\end{center}
    \vspace{-4mm}
    \caption{\textbf{Comparison of manifold-based interpolations of Gaussians.} The ground truth images and the estimated images by the neural ODE model when trained with different regularizers: a penalty on the time derivative of latent vectors, a penalty on the time derivative of decoded images, or our OT regularization. During training, the models only see images subsampled at an interval of 5 time points. OT regularization is the only regularizer maintaining consistent Gaussian shapes across time.
    }
    \label{fig:man_interp_squares}
\end{figure}

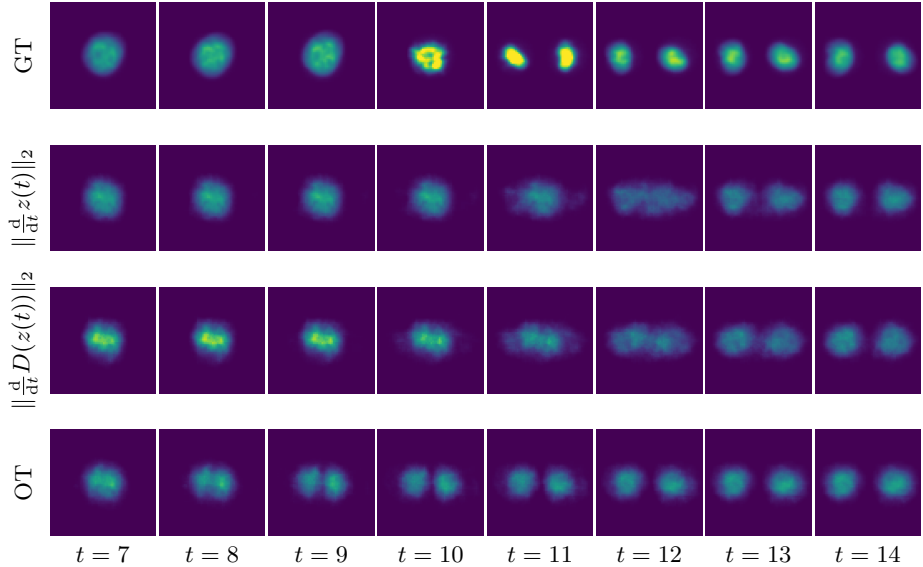
\begin{figure}[!t]
    \begin{center}
    \begin{tikzpicture}[
 image/.style = {text width=\myFigureScale\textwidth, 
                 inner sep=0pt, outer sep=0pt},
node distance = \myrowsep mm and \mycolumnsep mm
                        ] 

\matrix[column sep=\mycolumnsep mm, row sep=3mm] (gt_matrix) {
    \node [image] (frame1_gt) {\includegraphics[width=\linewidth]{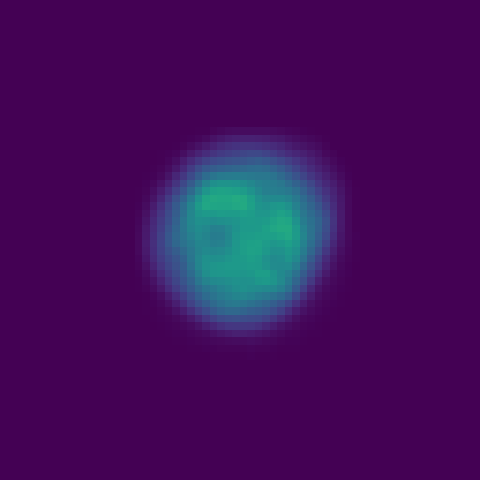}}; & 
    \node [image] {\includegraphics[width=\linewidth]{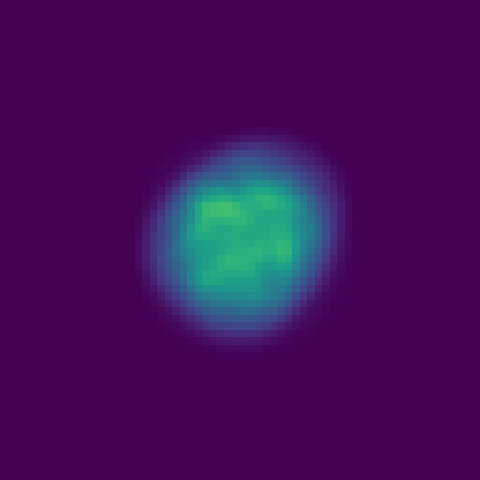}}; & 
    \node [image] {\includegraphics[width=\linewidth]{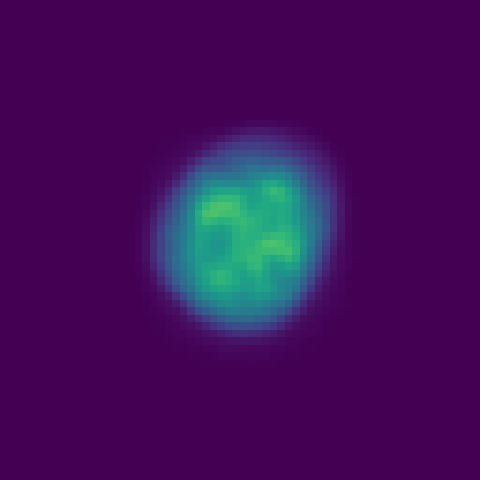}}; & 
    \node [image] {\includegraphics[width=\linewidth]{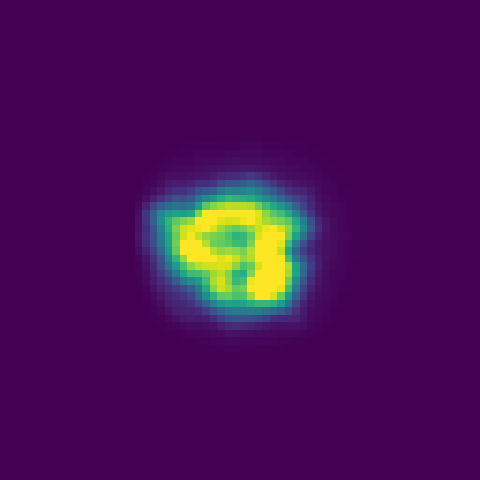}}; & 
    \node [image] {\includegraphics[width=\linewidth]{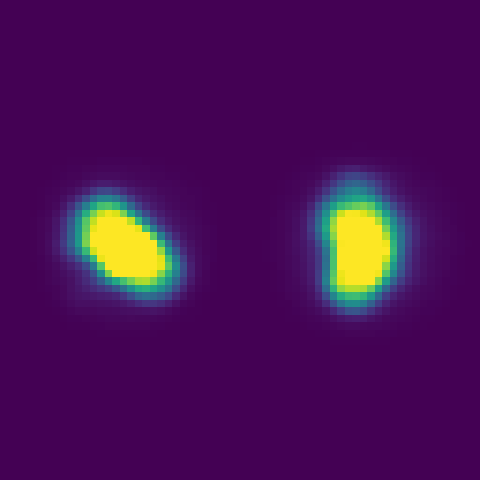}}; &
    \node [image] {\includegraphics[width=\linewidth]{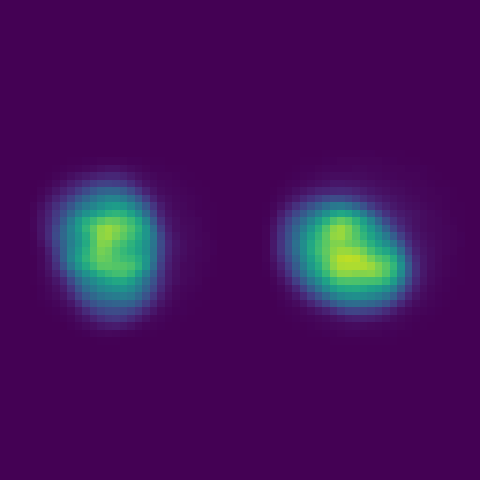}}; &
    \node [image] {\includegraphics[width=\linewidth]{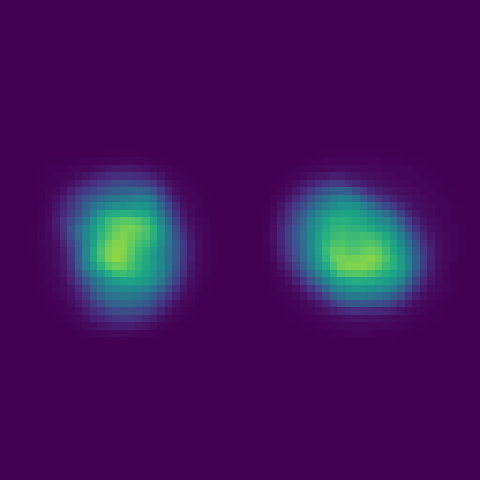}}; & 
    \node [image] {\includegraphics[width=\linewidth]{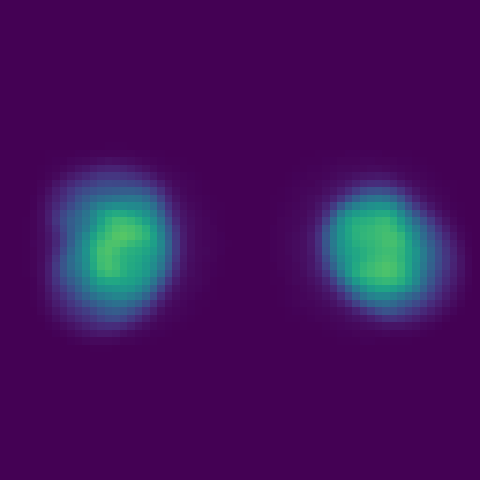}}; \\
};


\matrix[below=of gt_matrix, column sep=\mycolumnsep mm, row sep=3mm] (matrix_dzdt) {
    \node [image] (frame1_dzdt) {\includegraphics[width=\linewidth]{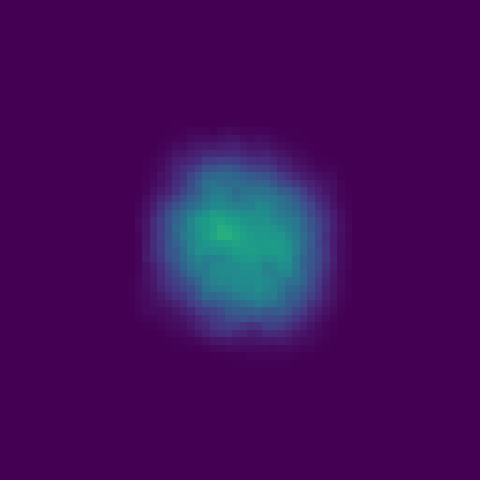}}; & 
    \node [image] {\includegraphics[width=\linewidth]{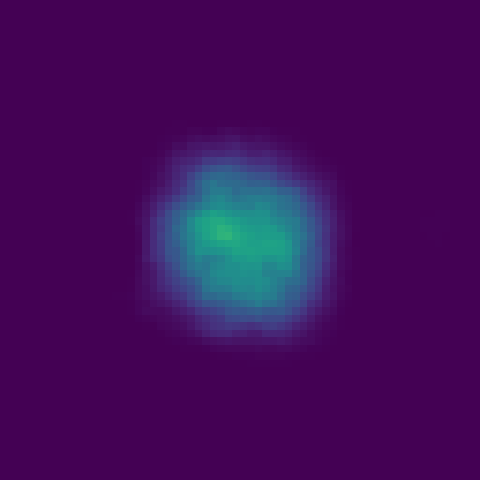}}; & 
    \node [image] {\includegraphics[width=\linewidth]{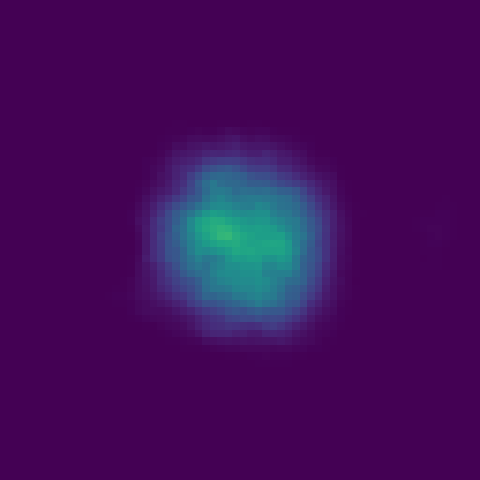}}; & 
    \node [image] {\includegraphics[width=\linewidth]{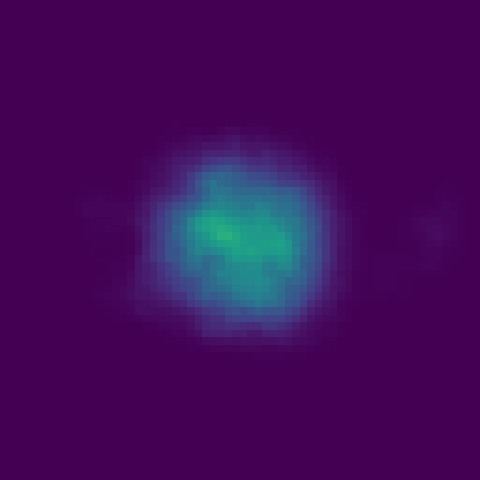}}; &
    \node [image] {\includegraphics[width=\linewidth]{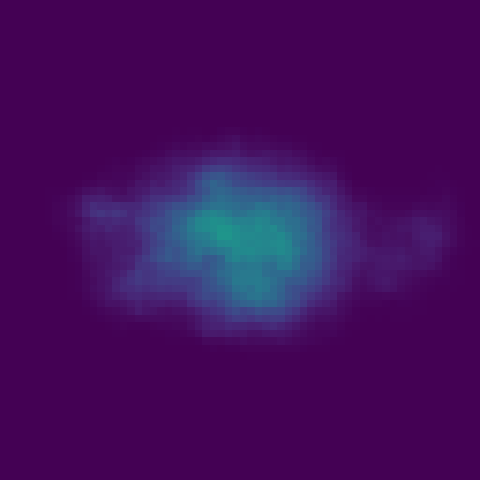}}; &
    \node [image] {\includegraphics[width=\linewidth]{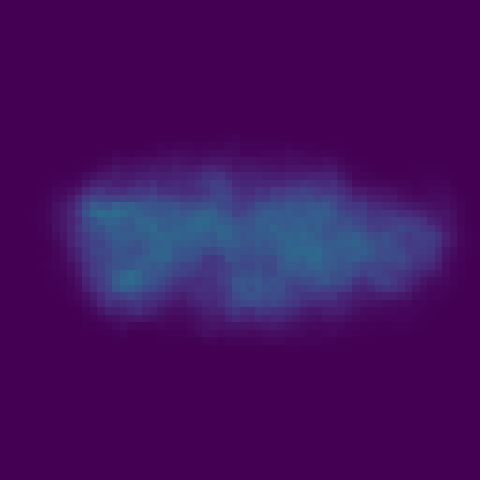}}; & 
    \node [image] {\includegraphics[width=\linewidth]{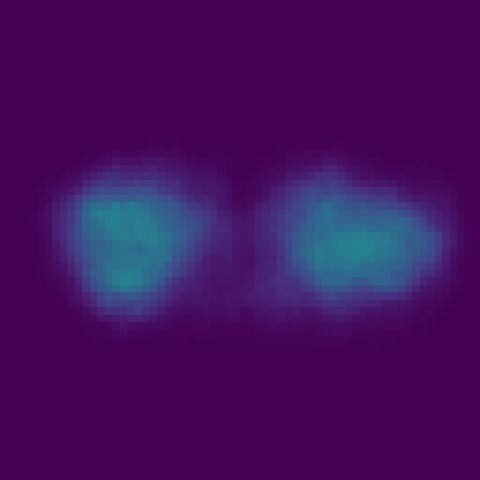}}; & 
    \node [image] {\includegraphics[width=\linewidth]{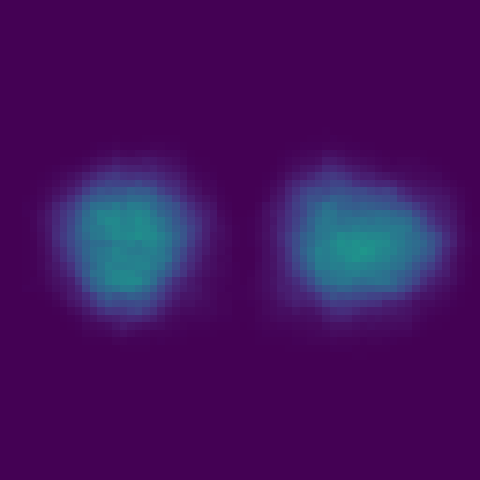}}; \\
};

\matrix[below=of matrix_dzdt, column sep=\mycolumnsep mm, row sep=3mm] (matrix_dIdt){
    \node [image] (frame1_dIdt) {\includegraphics[width=\linewidth]{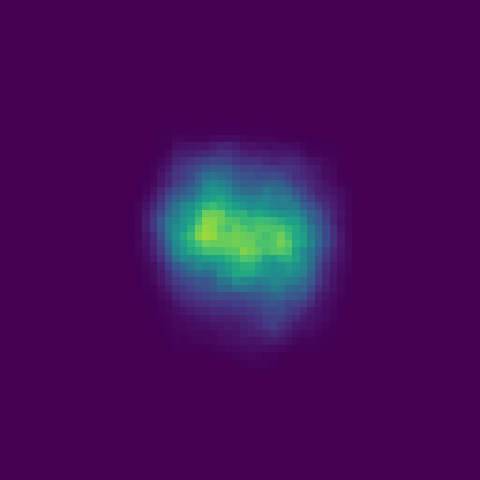}}; & 
    \node [image] {\includegraphics[width=\linewidth]{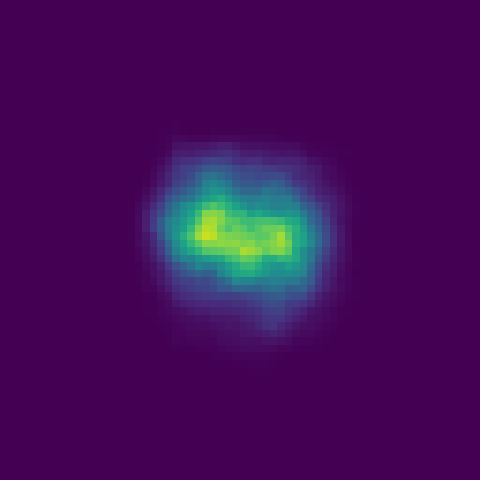}}; & 
    \node [image] {\includegraphics[width=\linewidth]{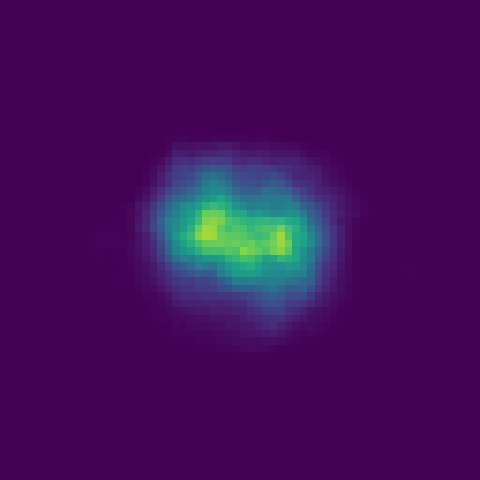}}; & 
    \node [image] {\includegraphics[width=\linewidth]{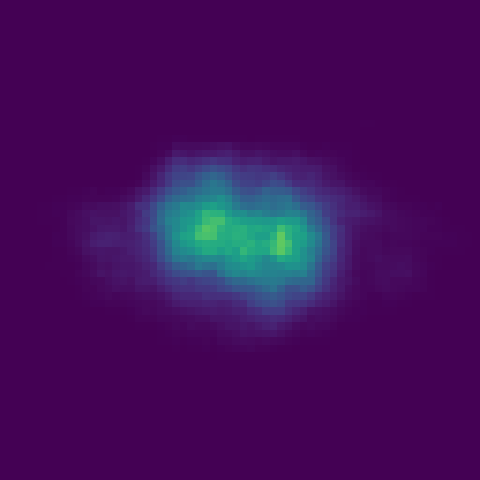}}; &
    \node [image] {\includegraphics[width=\linewidth]{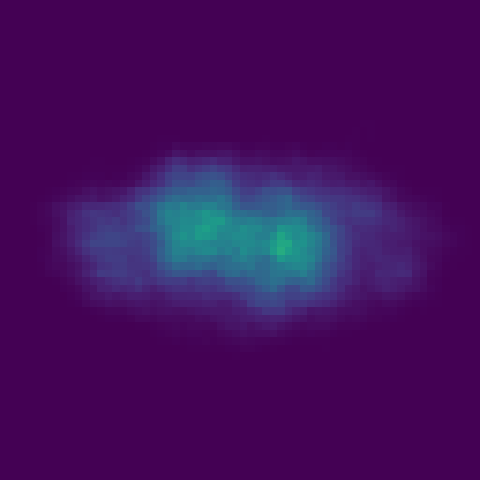}}; &
    \node [image] {\includegraphics[width=\linewidth]{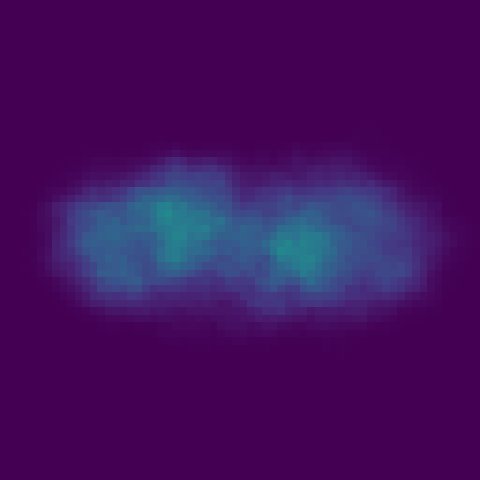}}; & 
    \node [image] {\includegraphics[width=\linewidth]{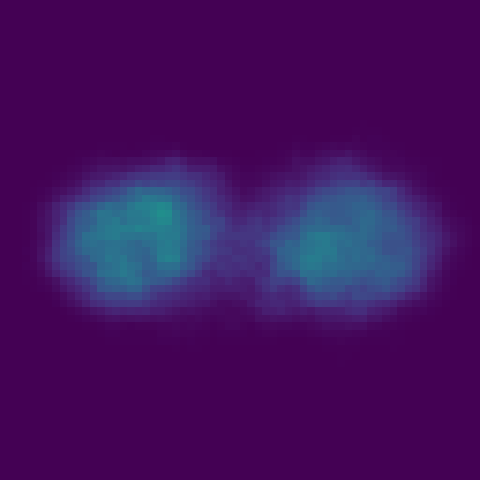}}; & 
    \node [image] {\includegraphics[width=\linewidth]{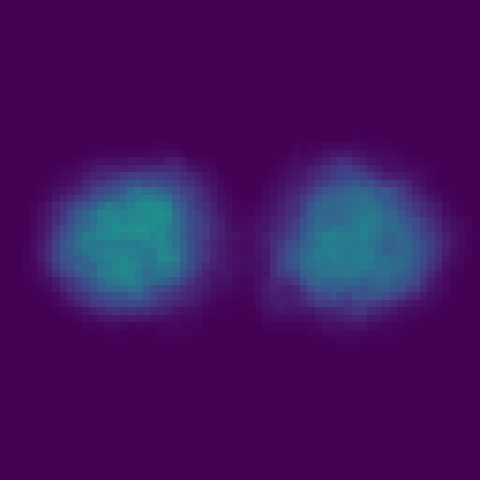}}; \\ 
};

\matrix[below=of matrix_dIdt, column sep=\mycolumnsep mm, row sep=3mm] (matrix_OT) {
    \node [image] (frame1) {\includegraphics[width=\linewidth]{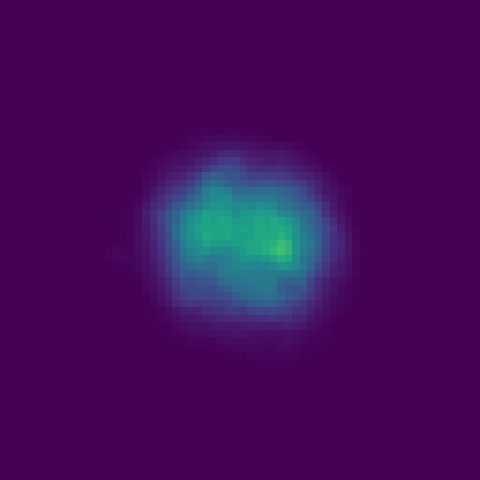}}; & 
    \node [image] (frame2) {\includegraphics[width=\linewidth]{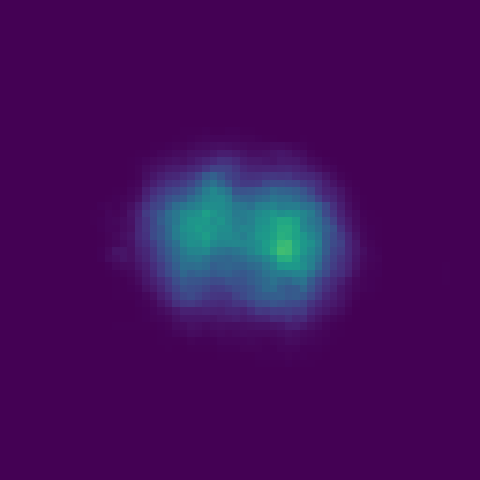}}; & 
    \node [image] (frame3) {\includegraphics[width=\linewidth]{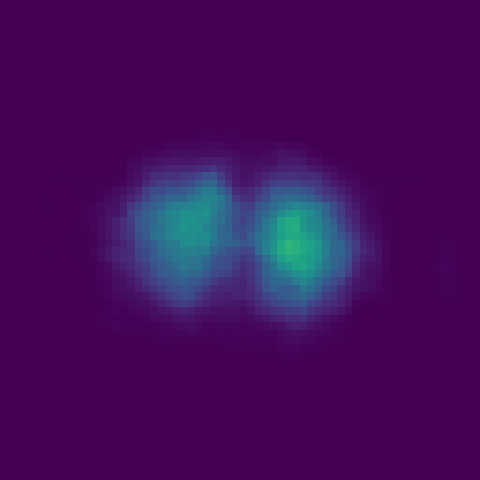}}; & 
    \node [image] (frame4) {\includegraphics[width=\linewidth]{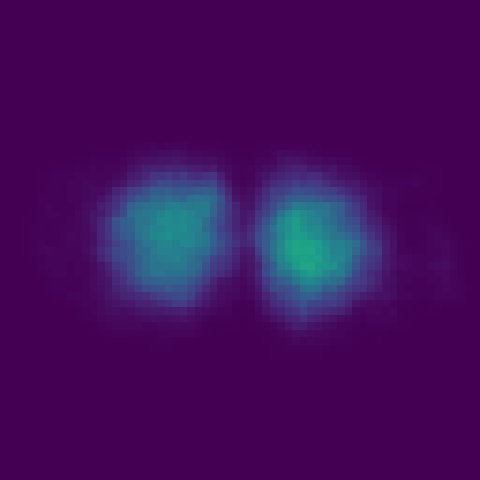}}; &
    \node [image] (frame5) {\includegraphics[width=\linewidth]{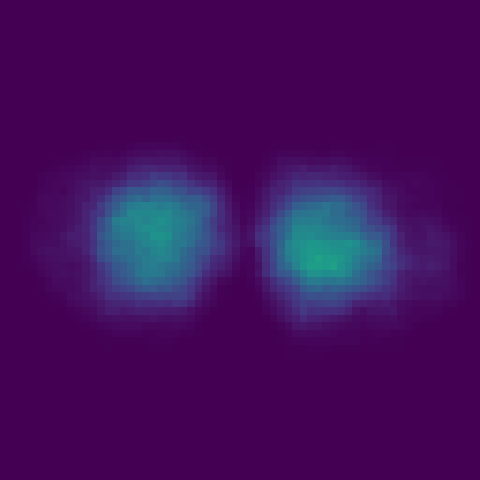}}; &
    \node [image] (frame6) {\includegraphics[width=\linewidth]{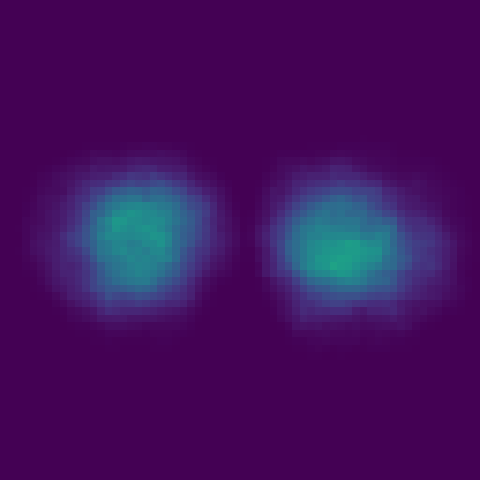}}; & 
    \node [image] (frame7) {\includegraphics[width=\linewidth]{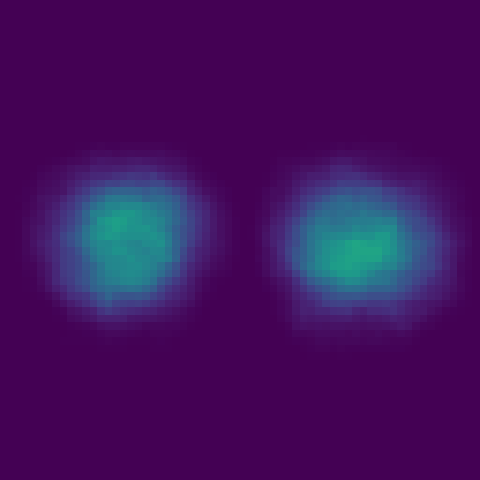}}; & 
    \node [image] (frame8) {\includegraphics[width=\linewidth]{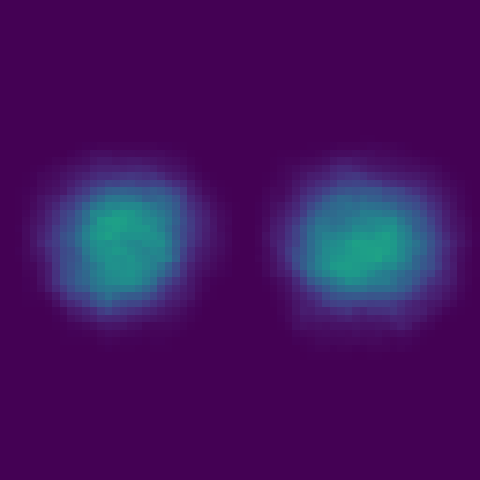}}; \\ 
};


\node [left=of frame1_dIdt, rotate=90, anchor=south] (dI_dt) {$\norm{\frac{\mathrm{d}}{ \mathrm{d}t}D(z(t))}_2$};
\draw (dI_dt |- gt_matrix) node[rotate=90] {GT};
\draw (dI_dt |- matrix_dzdt) node[rotate=90] {$\norm{\frac{\mathrm{d}}{\mathrm{d}t}z(t)}_2$};
\draw (dI_dt |- matrix_OT) node[rotate=90] {OT};

\node at ($(frame1.south) + (0.0,-0.25)$) {$t=7$};
\node at ($(frame2.south) + (0.0,-0.25)$) {$t=8$};
\node at ($(frame3.south) + (0.0,-0.25)$) {$t=9$};
\node at ($(frame4.south) + (0.0,-0.25)$) {$t=10$};
\node at ($(frame5.south) + (0.0,-0.25)$) {$t=11$};
\node at ($(frame6.south) + (0.0,-0.25)$) {$t=12$};
\node at ($(frame7.south) + (0.0,-0.25)$) {$t=13$};
\node at ($(frame8.south) + (0.0,-0.25)$) {$t=14$};

\end{tikzpicture}
\end{center}
    \vspace{-5mm}
    \caption{\textbf{Comparison of manifold-based interpolations of HeLa cells.} The ground truth images and the estimated images by the neural ODE model when trained with different regularizers: a penalty on the time derivative of latent vectors, a penalty on the time derivative of decoded images, or our OT regularization. During training, the models only see images subsampled at an interval of 5 time points. Only OT regularization obtains a clear and smooth cell division.
    }
    \label{fig:man_interp_HeLa}
\end{figure}

Figures \ref{fig:man_interp_squares} and \ref{fig:man_interp_HeLa} show images at locations where a Gaussian performs a quick jump and a cell divides, respectively. The interpolations with the latent space model result from the neural ODE, which uses the latent code of the initial image in the time series as the initial condition. 

All methods reconstruct the presented initial and final frames reasonably well, but the intermediate frames differ. In contrast to the other regularizations, OT regularization preserves consistent Gaussian shapes. For the HeLa cells in Figure~\ref{fig:man_interp_HeLa}, OT regularization smoothly separates the single cell into two. The other regularizers create a smeared transition, reducing intensity in the middle rather than distinctly splitting the cell. These results highlight that the OT prior can help manifold learning preserve the data's structure.

\subsection{Static and Dynamic Image Reconstruction}

To quantitatively assess the OT regularizer's impact, we explore its effect on static and dynamic reconstructions. Static reconstruction involves direct encoding and decoding, while dynamic reconstruction uses latent codes from the neural ODE model. 

Tables \ref{tab:static_recon_metrics} and \ref{tab:prognosis_recon_metrics} summarize the performance of static and dynamic reconstruction, respectively. For the static reconstruction of the Gaussian dataset, the OT prior preserves intensity and enforces shape awareness, leading to slightly improved reconstruction metrics compared to the other regularizers. In the more challenging HeLa datasets, all regularizers show similar performance in the static reconstruction.

\begin{table}[!b]
    \caption{\textbf{Static reconstruction evaluation.} We evaluate the models trained with various regularization types by encoding-decoding images and calculating the average value of the mean squared error (MSE) and average and standard deviation (between brackets) of the structural similarity index (SSIM). The best values are in bold.}
\begin{center}
	\begin{tabular}{c||rr|rr|rr|rr}
		\toprule
		\multicolumn{1}{c@{\quad}}{\textbf{\textit{Regularizer}}} & \multicolumn{8}{c}{\textbf{\textit{Dataset (metric)}}}                 \\
		\cmidrule(r){1-1} \cmidrule{2-9}
 & \multicolumn{2}{c|}{\textit{Gaussian (MSE)}} & \multicolumn{2}{c|}{\textit{Gaussian (SSIM)}} & \multicolumn{2}{c|}{\textit{HeLa (MSE)}} & \multicolumn{2}{c}{\textit{HeLa (SSIM)}} \\
\cmidrule{1-9} \morecmidrules \cmidrule{1-9}
  $\norm{\frac{\mathrm{d}}{\mathrm{d}t}z(t))}_2$ & $2.24\times 10^{-3}$ & & $0.956$ & $(0.040)$ & $1.41\times 10^{-3}$ & & $\textbf{0.907}$ & $(0.065)$ \\
  $\norm{\frac{\mathrm{d}}{\mathrm{d}t}D(z(t))}_2$ &  $1.96\times 10^{-3}$ & & $0.957$ & $(0.038)$ & $1.38\times 10^{-3}$ & & $\textbf{0.905}$ & $(0.067)$\\
    OT & $\textbf{1.66}\times 10^{-3}$ &  & $\textbf{0.960}$ & $(0.029)$ &  $\textbf{1.26}\times 10^{-3}$ & & $\textbf{0.905}$ & $(0.061)$\\
		\bottomrule
	\end{tabular}\label{tab:static_recon_metrics}
\end{center}
\end{table}

\begin{table}[tb]
    \caption{\textbf{Dynamic reconstruction evaluation.} We evaluate the models trained with various regularization types by using our neural ODE model for obtaining reconstructions and calculating the average value of the mean squared error (MSE) and average and standard deviation (between brackets) of the structural similarity index (SSIM). The best values are in bold.}
\begin{center}
	\begin{tabular}{c||rr|rr|rr|rr}
		\toprule
		\multicolumn{1}{c@{\quad}}{\textbf{\textit{Regularizer}}} & \multicolumn{8}{c}{\textbf{\textit{Dataset (metric)}}}                 \\
		\cmidrule(r){1-1} \cmidrule{2-9}
 & \multicolumn{2}{c|}{\textit{Gaussian (MSE)}} & \multicolumn{2}{c|}{\textit{Gaussian (SSIM)}} & \multicolumn{2}{c|}{\textit{HeLa (MSE)}} & \multicolumn{2}{c}{\textit{HeLa (SSIM)}} \\
\cmidrule{1-9} \morecmidrules \cmidrule{1-9}
  $\norm{\frac{\mathrm{d}}{\mathrm{d}t}z(t))}_2$ & $1.67\times 10^{-3}$ & & $0.961$ & $(0.030)$ & $9.14\times 10^{-3}$ & & $\textbf{0.757}$ & $(0.191)$ \\
  $\norm{\frac{\mathrm{d}}{\mathrm{d}t}D(z(t))}_2$ & $1.57\times 10^{-3}$ & & $0.959$ & $(0.033)$ & $8.74\times 10^{-3}$ & & $\textbf{0.762}$ & $(0.176)$\\
    OT & $\textbf{0.98}\times 10^{-3}$ &  & $\textbf{0.966}$ & $(0.019)$ & $\textbf{8.31}\times 10^{-3}$ & & $\textbf{0.773}$ & $(0.189)$  \\
		\bottomrule
	\end{tabular}\label{tab:prognosis_recon_metrics}
\end{center}
\end{table}

Comparing Table \ref{tab:static_recon_metrics} and \ref{tab:prognosis_recon_metrics} shows that for each regularizer, dynamic reconstruction outperforms static reconstruction on the Gaussian dataset but underperforms on the HeLa dataset. This difference mainly stems from errors in the latent prediction process rather than the decoding step. In particular, dynamic reconstruction performs well on the Gaussian dataset due to the dataset's relatively simple temporal behavior. Combined with the fact that we emphasize dynamic reconstruction more than static reconstruction during training, this results in improved reconstruction metrics compared to the static case. In contrast, predicting future states from an initial image of the HeLa dataset is significantly more challenging, especially due to its heterogeneous intracellular motion. As a result, inaccuracies in latent prediction primarily drive reconstruction errors, resulting in worse reconstruction metric values.

\subsection{Interpolations Consistent with Manifold Structure}
This section investigates the benefits of manifold learning compared to optimal transport interpolations. Specifically, we perform a Euclidean and Wasserstein interpolation between two time points and compare it with the trajectory predicted by our neural ODE model. This comparison sheds light on the learned structure of the data manifold. The idea is that an Euclidean or Wasserstein interpolation does not consider this structure and, hence, does not follow the structure of the image manifold. In contrast, the neural ODE model offers an alternative path that more faithfully aligns with the dynamic data and its underlying manifold structure. In other words, while Euclidean and Wasserstein interpolations are the shortest paths not consistent with the image manifold and the temporal dynamics of the data, our model learns the shortest paths consistent with the geometry of the data manifold and the temporal behavior of the images. 

\begin{figure}[!t]
    \centering
    \begin{tikzpicture}[
 image/.style = {text width=\myFigureScale\textwidth, 
                 inner sep=0pt, outer sep=0pt},
node distance = \myrowsep mm and \mycolumnsep mm
                        ] 



\matrix[column sep=\mycolumnsep mm, row sep=3mm] (gt_matrix) {
    \node [image] {\includegraphics[width=\linewidth]{figures_revision/gt_hela/track288_7.png}}; & 
    \node [image] {\includegraphics[width=\linewidth]{figures_revision/gt_hela/track288_8.png}}; & 
    \node [image] {\includegraphics[width=\linewidth]{figures_revision/gt_hela/track288_9.png}}; & 
    \node [image] {\includegraphics[width=\linewidth]{figures_revision/gt_hela/track288_10.png}}; & 
    \node [image] {\includegraphics[width=\linewidth]{figures_revision/gt_hela/track288_11.png}}; &
    \node [image] {\includegraphics[width=\linewidth]{figures_revision/gt_hela/track288_12.png}}; &
    \node [image] {\includegraphics[width=\linewidth]{figures_revision/gt_hela/track288_13.png}}; & 
    \node [image] {\includegraphics[width=\linewidth]{figures_revision/gt_hela/track288_14.png}}; \\
};


\matrix[below=of gt_matrix, column sep=\mycolumnsep mm, row sep=3mm] (matrix_dzdt) {
    \node [image] {\includegraphics[width=\linewidth]{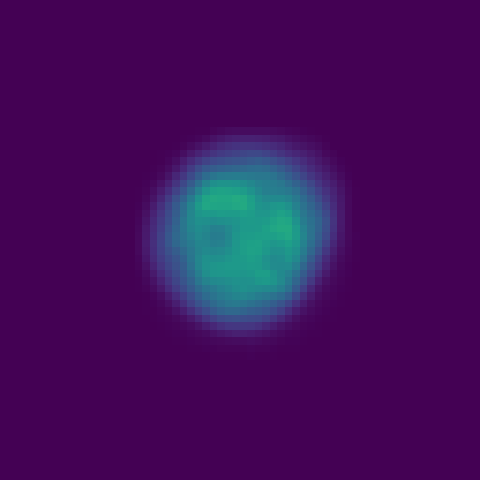}}; & 
    \node [image] {\includegraphics[width=\linewidth]{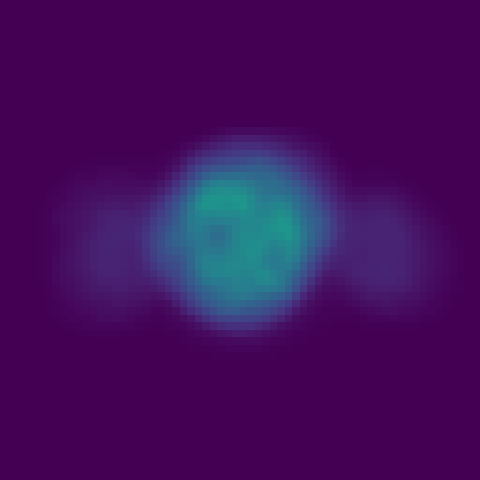}}; & 
    \node [image] {\includegraphics[width=\linewidth]{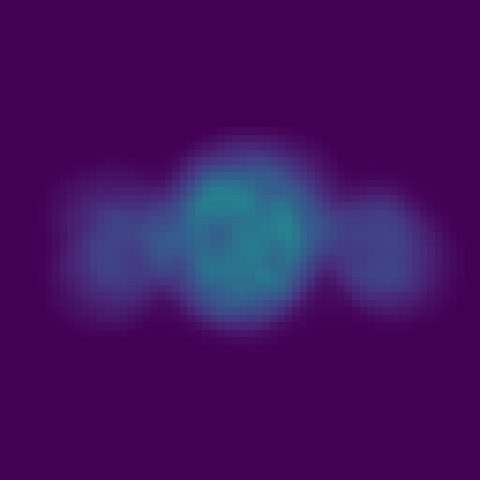}}; &
    \node [image] {\includegraphics[width=\linewidth]{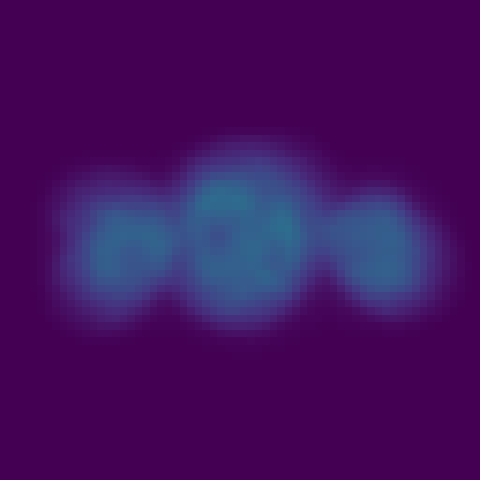}}; &
    \node [image] {\includegraphics[width=\linewidth]{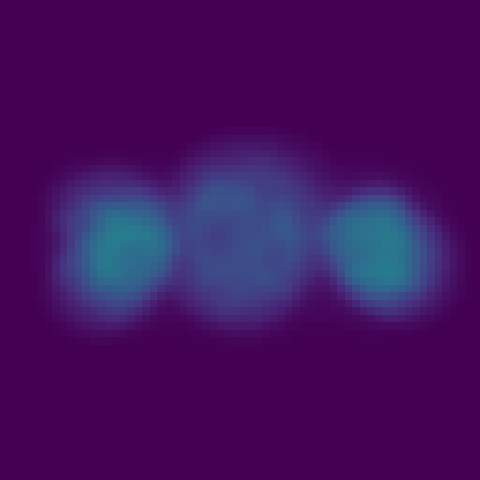}}; & 
    \node [image] {\includegraphics[width=\linewidth]{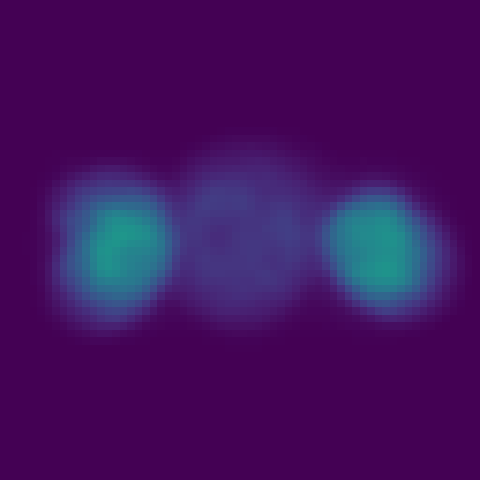}}; & 
    \node [image] {\includegraphics[width=\linewidth]{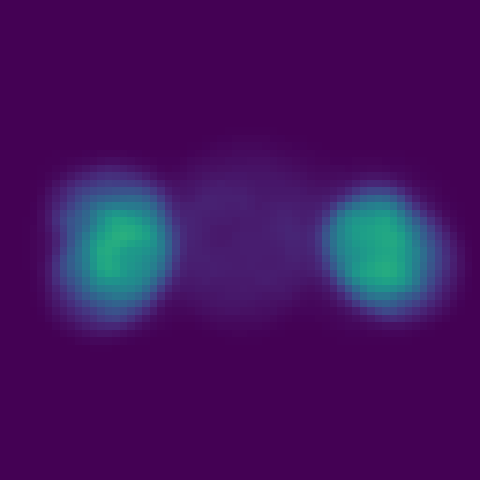}}; & 
    \node [image] {\includegraphics[width=\linewidth]{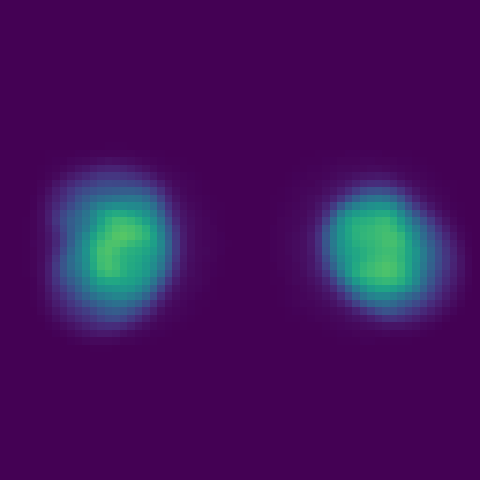}}; \\
};


\matrix[below=of matrix_dzdt, column sep=\mycolumnsep mm, row sep=3mm] (matrix_dIdt){
    \node [image] {\includegraphics[width=\linewidth]{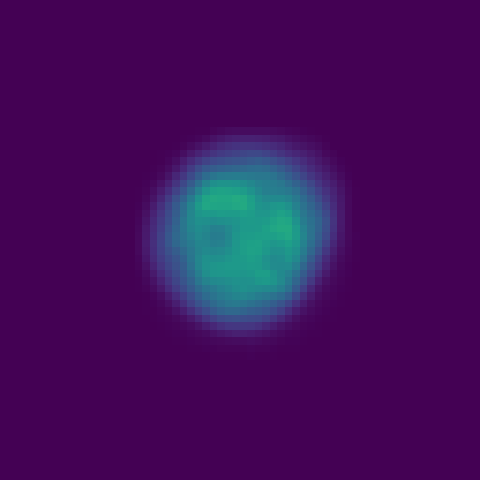}}; & 
    \node [image] {\includegraphics[width=\linewidth]{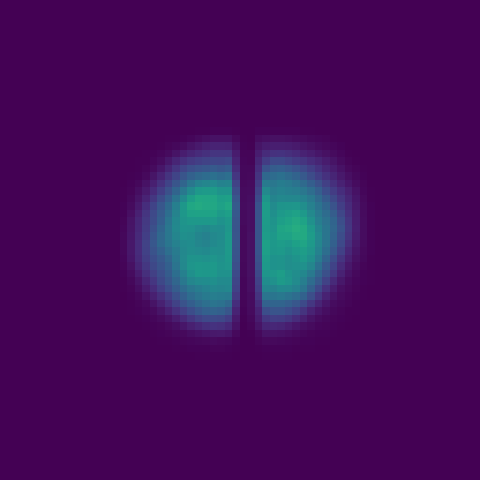}}; & 
    \node [image] {\includegraphics[width=\linewidth]{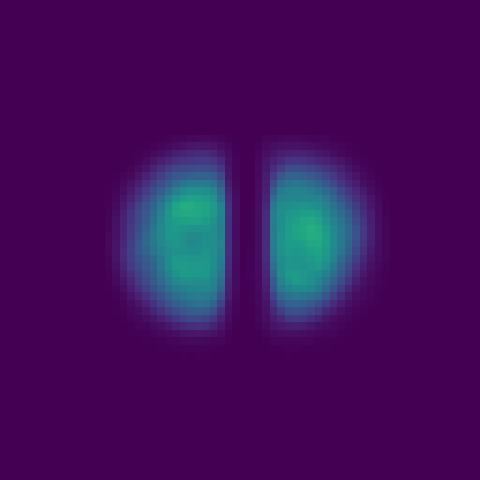}}; &
    \node [image] {\includegraphics[width=\linewidth]{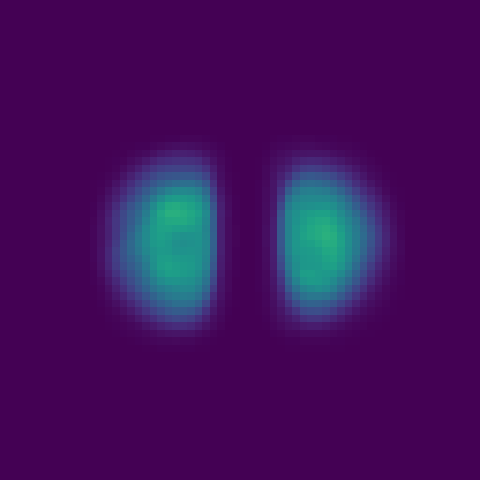}}; &
    \node [image] {\includegraphics[width=\linewidth]{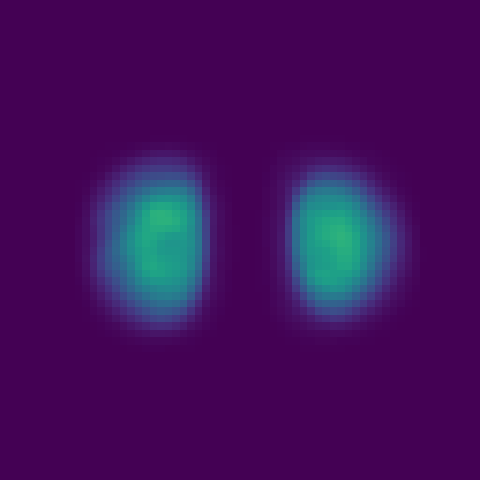}}; & 
    \node [image] {\includegraphics[width=\linewidth]{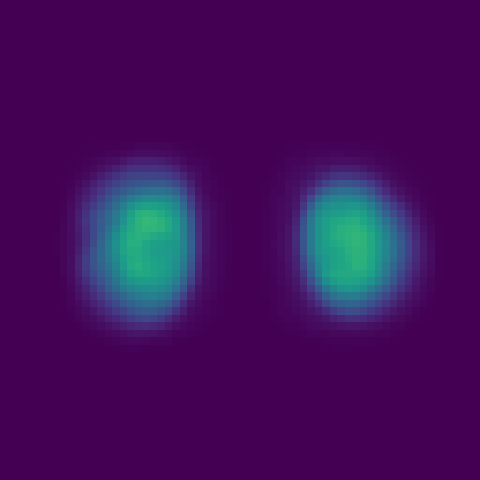}}; & 
    \node [image] {\includegraphics[width=\linewidth]{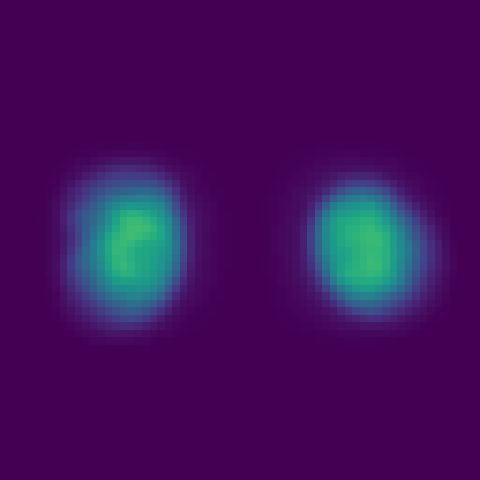}}; & 
    \node [image] {\includegraphics[width=\linewidth]{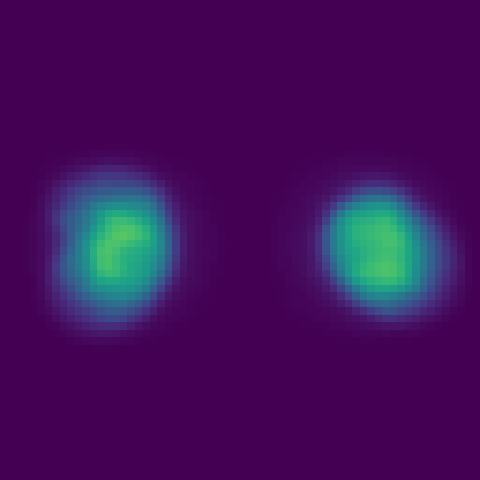}}; \\
};

\matrix[below=of matrix_dIdt, column sep=\mycolumnsep mm, row sep=3mm] (matrix_OT) {
    \node [image] (frame1) {\includegraphics[width=\linewidth]{figures_revision/w2_hela/recon_7.png}}; & 
    \node [image] (frame2) {\includegraphics[width=\linewidth]{figures_revision/w2_hela/recon_8.png}}; & 
    \node [image] (frame3) {\includegraphics[width=\linewidth]{figures_revision/w2_hela/recon_9.png}}; & 
    \node [image] (frame4) {\includegraphics[width=\linewidth]{figures_revision/w2_hela/recon_10.png}}; &
    \node [image] (frame5) {\includegraphics[width=\linewidth]{figures_revision/w2_hela/recon_11.png}}; &
    \node [image] (frame6) {\includegraphics[width=\linewidth]{figures_revision/w2_hela/recon_12.png}}; & 
    \node [image] (frame7) {\includegraphics[width=\linewidth]{figures_revision/w2_hela/recon_13.png}}; & 
    \node [image] (frame8) {\includegraphics[width=\linewidth]{figures_revision/w2_hela/recon_14.png}}; \\
};

\node [left=of frame1_dIdt, rotate=90, anchor=south] (dI_dt) {\small $\mathcal{W}_2$};
\draw (dI_dt |- gt_matrix) node[rotate=90] {GT};
\draw (dI_dt |- matrix_dzdt) node[rotate=90] {$l_2$};
\draw (dI_dt |- matrix_OT) node[rotate=90] {Manifold};

\node at ($(frame1.south) + (0.0,-0.25)$) {$t=7$};
\node at ($(frame2.south) + (0.0,-0.25)$) {$t=8$};
\node at ($(frame3.south) + (0.0,-0.25)$) {$t=9$};
\node at ($(frame4.south) + (0.0,-0.25)$) {$t=10$};
\node at ($(frame5.south) + (0.0,-0.25)$) {$t=11$};
\node at ($(frame6.south) + (0.0,-0.25)$) {$t=12$};
\node at ($(frame7.south) + (0.0,-0.25)$) {$t=13$};
\node at ($(frame8.south) + (0.0,-0.25)$) {$t=14$};

\end{tikzpicture}
    \caption{\textbf{Barycentric vs. Manifold interpolation.} The ground truth images, the estimated images by $l_2$ or Wasserstein $\mathcal{W}_2$ interpolation, and the estimated images by the neural ODE manifold model. $l_2$ interpolation does not resemble a (divided) cell, $\mathcal{W}_2$ splits the cells prematurely, and manifold interpolation avoids both issues.}
    \label{fig:comparison_to_plain_interpolation}
\end{figure}

The results on the HeLa dataset, shown in Figure \ref{fig:comparison_to_plain_interpolation}, demonstrate that the neural ODE model effectively captures the dynamics of the data while respecting the image manifold. It accurately follows the structure of cell division, and the generated images resemble the ground truth divided cells. On the other hand, the Euclidean interpolation fails to account for the manifold structure and the temporal evolution of the data, resulting in images that neither properly resemble a divided cell nor represent the underlying dynamics. The Wasserstein interpolations better align with the image manifold. However, at the early time points, the cells appear as if cut in half, which deviates from the smoother shapes seen in realistic cells. Furthermore, the cell division occurs too early compared to the ground truth. 

These observations suggest that manifold learning can help align the non-learned methods discussed in Section \ref{sec:rel_work_dyn_img} with the geometry of the data manifold and the temporal behavior of the data.

\section{Conclusion and Future Work}
This work explores the effects of combining a low-dimensional manifold assumption with a temporal OT prior, specifically in the context of dynamic imaging. We designed an autoencoder model where movement in latent space amounts to movement in image space. We regularize the resulting manifold of static images by combining a neural ordinary differential equation, dynamic image data, and the dynamic interpretation of OT. Our results demonstrate that the OT prior enhances manifold learning approaches, while the manifold improves OT by incorporating data structure.

While this work focuses on balanced OT, the images we deal with have varying masses. The varying mass highlights the relevance of unbalanced OT, which deals with objects with different masses, making it an interesting direction for future research. Additionally, our experiments do not consider scenarios involving a forward operator. Future work could explore these problems and extend the investigation of the interplay between OT and manifold learning.

\begin{credits}
\subsubsection{\discintname}
The authors have no competing interests to declare that are
relevant to the content of this article.
\end{credits}
%
%
\bibliographystyle{splncs04}
\bibliography{references}
%




\end{document}